# Properties of the $a_1$ Meson from Lattice QCD

Matthew Wingate, Thomas DeGrand

*Department of Physics, University of Colorado, Boulder, CO 80309, USA*

Sara Collins, Urs M. Heller

*Supercomputer Computations Research Institute, Florida State University, Tallahassee, FL 32306, USA*

(April 19, 1995)

## Abstract

We determine the mass and decay constant of the $a_1$ meson using Monte Carlo simulation of lattice QCD. We find $M_{a_1} = 1250 \pm 80$ MeV and $f_{a_1} = 0.30 \pm 0.03$ (GeV)$^2$, in good agreement with experiment.

First seen as a resonance in pion-proton scattering, the axial vector meson $a_1$ remained an elusive prey due to its large width and the presence of strong background signals [1]. Since 1986 the properties of the $a_1$ have been measured more precisely through the decay of the tau lepton to three pions: $\tau \to \nu_\tau a_1 \to \nu_\tau \rho \pi \to \nu_\tau \pi \pi \pi$ [2].

The decay of the $a_1$ is parameterized by the $W \to a_1$ vertex which, in the conventions of Tsai [3], is defined in terms of a dimensionful decay constant $f_{a_1}$ as

$$\Gamma^{\mu\nu}(W \to a_1) \equiv -i\, f_{a_1}(p^2)\, g^{\mu\nu}. \tag{1}$$

In this definition $f_{a_1}$ has units of $[\text{mass}]^2$. The mass and decay constant are both quantities accessible to measurement from lattice QCD. In this work we describe a lattice calculation of the mass and decay constant of the $a_1$ meson and compare our results to experiment.

The quantities measured on the lattice which give the $a_1$ mass and decay constant are two point correlation functions

$$C_{ij}(t) = \langle 0|O_i(t)O_j(0)|0\rangle, \tag{2}$$

which, for large $t$, reduce to a single decaying exponential (plus boundary terms):

$$C(t) = \frac{Z_i Z_j}{2aM}\, e^{-aMt}, \tag{3}$$

where $M$ is the mass of the lightest particle that couples to $O$, $a$ is the lattice spacing, and the $Z$'s are the appropriate source and sink matrix elements, $Z_i = \langle 0|O_i|M\rangle$. In order to couple to the $a_1$, the operator we use is the local axial-vector current $A_i = \bar\psi \gamma_i \gamma_5 \psi$. For the operator which creates the state, we gauge fix to Coulomb gauge and use an axial current operator in which the quark and antiquark are created with Gaussian spatial distributions about a source point. To measure the decay constant we compute two kinds of correlation functions, ones in which the second operator is identical to the source and ones in which the second operator is the local current. A simultaneous fit to the two correlators allows an extraction of the mass and decay constant. We parameterize the matrix element for the local sinks as

$$Z_p = \langle 0|A_i|a_1\rangle \equiv M_{a_1}^2 f_{a_1}^{Lat}\epsilon_i, \tag{4}$$

where $f_{a_1}^{Lat}$ is dimensionless.

The simulations were carried out on the Connection Machine CM-2 at the Supercomputer Computations Research Institute at Florida State University. We used the ensemble of configurations generated by the HEMCGC collaboration with two flavors of dynamical staggered quarks [4]. The configurations were generated using the hybrid molecular dynamics (HMD) algorithm [5]. The size of the lattices is $16^3 \times 32$, the lattice coupling is $\beta = 6/g^2 = 5.6$, and the dynamical quark masses in lattice units $a$ are $am_q = 0.010$ and 0.025. Periodic boundary conditions were used in all four directions of the lattice. The total simulation length was 2000 simulation time units (with the normalization of ref. [6]) at each quark mass value. We analyzed lattices spaced by 20 HMD time units, for a total of 100 lattices at each mass value.

The spectroscopy was computed with six values of the Wilson quark hopping parameter: $\kappa = 0.1600, 0.1585, 0.1565, 0.1525, 0.1410,$ and $0.1320$. The first three values are rather light quarks (the pseudoscalar mass in lattice units ranges from about 0.25 to 0.45), and the other three values correspond to heavy quarks (pseudoscalar mass from 0.65 to 1.5). Our inversion technique is conjugate gradient with preconditioning via incomplete lower-upper (ILU) decomposition by checkerboards [7]. For more details about the dynamical staggered fermion simulations see ref. [4,8]. Since we use sources for the propagators which are extended in space, we fix gauge to lattice Coulomb gauge using an overrelaxation method [9].

A few words are in order about our implicit assumptions. The state created by the correlator at asymptotic Euclidean time $t$ is the on-shell (physical) $a_1$, and so the matrix element calculated is the on-shell decay constant. In tau decay the produced $a_1$ is usually virtual (not at the peak of its resonance) and therefore is not "on shell". In practice the off-shell production of the $a_1$ is entangled with its decay into three pions, which has to be modeled somehow. To make contact with tau decay, we follow the phenomenological analysis of Isgur, Morningstar, and Reader (IMR) [10] which parameterizes tau decay into $a_1$ in terms of the on-shell decay constant. The calculation is in complete analogy with the decay chain $e^+e^- \to \rho \to \pi\pi$, where the rho decay constant is taken to be a constant $f_\rho$ and the decay into pions is modeled by an on-shell rho decay matrix element. This calculation neglects processes where the $a_1$ flows backwards in time. In addition to extracting $f_{a_1}$ from data, IMR also present a calculation of it in a quark model; our work replaces the quark model calculation with a direct determination of the decay constant from QCD.

Presently it is not possible in lattice QCD to include the decay of the $a_1$ into $\rho$ and $\pi$ mesons, which would give us the off-shell form factors.

In order to find the continuum $a_1$ mass $M_{a_1}$ and decay constant $f_{a_1}^{Cont}$, it is necessary to compute four quantities: the lattice mass $aM_{a_1}$, the lattice spacing $a$, the strong coupling constant defined in a particular prescription $\alpha_V$, and the lattice decay constant $f_{a_1}^{Lat}$.

The lattice mass is measured by performing a two-parameter correlated fit [11] to the axial-vector propagators for the three lightest quark-antiquark pairs. The results of these fits are presented elsewhere [12]. Since the valence quarks in the simulation are much heavier than the physical $u$ and $d$ quarks, we extrapolate linearly to zero quark mass. We find

$$aM_{a_1} = \begin{cases} 0.670 \pm 0.018 & \text{for } am_q = 0.010 \\ 0.738 \pm 0.049 & \text{for } am_q = 0.025. \end{cases} \quad (5)$$

The errors in these numbers are purely statistical.

One expects [13] the mass difference between heavy $Q\bar{Q}$ $L=1$ states and $L=0$ states to be weakly dependent upon the quark mass, and so we use the quarkonium S-P mass splitting to determine the lattice spacing. This calculation is performed in our companion work [12] with the conclusion that

$$a^{-1} = \begin{cases} 1900 \pm 50 \pm 100 \text{ MeV} & \text{for } am_q = 0.010 \\ 1660 \pm 110 \pm 100 \text{ MeV} & \text{for } am_q = 0.025. \end{cases} \quad (6)$$

The first error is statistical and the second is our estimate of the systematic uncertainties implicit in choosing the S-P mass splitting as the physical quantity to set the scale. Thus we find for the mass of the $a_1$

$$M_{a_1} = \begin{cases} 1270 \pm 80 \text{ MeV} & \text{for } am_q = 0.010 \\ 1230 \pm 140 \text{ MeV} & \text{for } am_q = 0.025. \end{cases} \quad (7)$$

Although the quantities we calculate on the lattice change with the sea quark lattice mass as seen in (5) and (6), the physical observables are independent of the mass of the dynamical fermions within errors. This has been true of all lattice calculations of the QCD spectrum to date. Therefore we treat the two simulations as independent and average the two results giving a physical $a_1$ mass of

$$M_{a_1} = 1250 \pm 80 \text{ MeV}. \quad (8)$$

This is in very good agreement with the experimental value 1230(40) MeV [14]. With these definitions of the lattice spacing the same simulations predict rho masses of 680(40) and 640(60) MeV and nucleon masses of 990(60) and 920(80) MeV for lattice quark masses $am_q = 0.010$ and 0.025, respectively.

To compute the lattice decay constant, we perform a three-parameter correlated fit to the two different correlation functions for the three lightest quarkonium states and extrapolate to zero quark mass. We find

$$f_{a_1}^{Lat} = \begin{cases} 0.881 \pm 0.036 & \text{for } am_q = 0.010 \\ 0.803 \pm 0.056 & \text{for } am_q = 0.025. \end{cases} \quad (9)$$

In order to extract the continuum value from the lattice value, we must set the scale with the lattice spacing and multiply by a renormalization factor to convert from lattice to continuum regularization. We use the tadpole-improved formalism of Ref. [15], where

$$f_{a_1}^{Cont} = f_{a_1}^{Lat} (aM_{a_1})^2 \left(\frac{1}{a}\right)^2 \frac{1}{4} (1 - 0.31\alpha_V), \quad (10)$$

and $\alpha_V$ is the running coupling constant defined through the potential on the lattice at a momentum scale $q = 1.03/a$. Through the expectation value of the plaquette, we measured the strong coupling at $q = 3.41/a$ and ran it down to $q = 1.03/a$ using the two-loop QCD beta function. For our two values of the sea quark mass, $am_q = 0.010$ and 0.025, we found that $\alpha_V = 0.29(5)$ and $0.31(5)$. Therefore, the continuum $a_1$ decay constant is calculated to be

$$f_{a_1}^{Cont} = \begin{cases} 0.33 \pm 0.04 \text{ (GeV)}^2 & \text{for } am_q = 0.010 \\ 0.27 \pm 0.05 \text{ (GeV)}^2 & \text{for } am_q = 0.025. \end{cases} \quad (11)$$

Again, we treat these results as independent determinations of the same quantity and average them to get

$$f_{a_1}^{Cont} = 0.30 \pm 0.03 \text{ (GeV)}^2. \quad (12)$$

The calculation of a branching fraction is very complicated due to the large width of the $a_1$ [10,2]. The value of $f_{a_1}$ quoted in the 1989 paper by Isgur, Morningstar, and Reader was derived phenomenologically using the absolute rate for $\tau \to \nu_\tau \pi \pi \pi$ [16]. They found that the experimental data at that time give the following decay constant [17]

$$f_{IMR}^{Exp} = 0.25 \pm 0.02 \text{ (GeV)}^2. \quad (13)$$

This is in good agreement with our theoretical calculation.

We can attempt a similar calculation of the mass and decay constant of the strange axial vector meson, the $K_1$. However, this calculation is not complete. The physical $K_1(1270)$ and $K_1(1400)$ mesons are combinations of the $^3P_1$ and $^1P_1$ states [19]. While the correlator in equations (2) and (3) couples asymptotically to the lightest state which has nonzero overlap with its operators, the contamination of the next state only dies away as $\exp(-\Delta M t)$, where $\Delta M$ is the mass gap. The two $K_1$ states are close in mass, and since we do not have multiple operators to do a multistate fit [20], we cannot be sure that we are seeing the lighter $K_1$. Nevertheless, in the absence of other theoretical predictions, we attempt a calculation.

Fixing the $\phi$ meson to its physical mass gives us $\kappa_s$, the Wilson hopping parameter corresponding to the the strange quark mass. Then we are able to find the lattice mass and lattice decay constant by extrapolating the lighter quark to zero mass while holding the heavy quark at the strange quark mass. Finally, we must correct the factor of 1/4 in eqn. (10) since only one quark is light [15,21]:

$$f_{K_1}^{Cont} = f_{K_1}^{Lat} (aM_{K_1})^2 \left(\frac{1}{a}\right)^2 \frac{1}{2} \sqrt{1 - \frac{3\kappa_s}{4\kappa_c}} (1 - 0.31\alpha_V), \tag{14}$$

where $\kappa_c$ is the hopping parameter where the renormalized quark masses vanish; *i.e.* $\kappa \to \kappa_c$ is the chiral limit.

With these caveats, we report lattice measurements of $M_{K_1} = 1390 \pm 80$ MeV and $f_{K_1} = 0.33 \pm 0.03$ (GeV)$^2$.

We conclude by remarking that these calculations are extremely straightforward extensions of commonly performed lattice measurements. We encourage other lattice groups to do them, so that the sensitivity of these predictions to the standard lattice systematics (lattice spacing, simulation volume, quenching vs. dynamical quark mass) can be explored. We also encourage the lattice community to consider tau decay as a source of phenomenological problems.

## ACKNOWLEDGMENTS

This work was supported by the U.S. Department of Energy under grants DE-FG05-85ER250000, DE-FG05-92ER40742, and DE-FG02–92ER–40672. Simulations were carried out at the Supercomputer Computations Research Institute at Florida State University. T.D. and M.W. would like to thank Jim Smith and Bill Ford for introducing them to this problem, and Claude Bernard for a helpful conversation.